\documentclass[9pt,article,twoside]{rmaa-rho-class/rmaa-rho}
\RMxAAtemplatetype{\RMxAA} 

\usepackage{hyperref}
\usepackage{arydshln}

\usepackage{caption}
\newcommand{\gband}{\ensuremath{\mathrm{g'}}}
\newcommand{\rband}{\ensuremath{\mathrm{r'}}}
\newcommand{\iband}{\ensuremath{\mathrm{i'}}}

\newcommand{\Bband}{\ensuremath{\mathrm{B}}}

\newcommand{\orcid}[1]{\textsuperscript{\href{http://orcid.org/#1}{\hskip2pt\includegraphics[width=8pt]{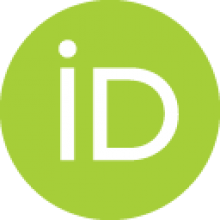}}}}

\vol{61}
\pages{30-36}
\thisyear{2025}
\doi{\href{https://www.astroscu.unam.mx/rmaa/RMxAA..XX-X}{https://www.astroscu.unam.mx/rmaa/RMxAA..XX-X}}

\title{A COLIBRI Photometric Study of SN 2025bvm: A Normal, Slowly Declining Type Ia Supernova}

\author[1]{D. H. Gonz\'{a}lez-Buitrago \orcidlink{0000-0002-9280-1184}}
\author[1]{Ma. T. Garc\'ia-D\'iaz \orcidlink{0000-0002-9772-5555}}
\author[1]{A. E. Montoya-Olivo} 
\author[1]{S. Sánchez-Sanjuán}
\author[2]{H. Avila-Mogollon}

\affil[1]{Universidad Nacional Autónoma de México, Instituto de Astronomía, AP 106,  Ensenada 22800, BC, México}
\affil[2]{Colegio Nuestra señora de la Concepción, Concepción Santander, Colombia}

\leadauthor{Gonz\'{a}lez-Buitrago et al.}
\smalltitle{LaTex Macro RMxAA}

\corres{Diego Gonz\'{a}lez-Buitrago}
\email{dgonzalez@astro.unam.mx}

\received{April 16th, 2024}
\accepted{\today}

\license{Texto de la licencia aquí}

\setbool{rho-abstract}{true} 
\setbool{rho-resumen}{true} 

\begin{abstract}
We present 121 days of multi-band (\Bband, \gband, \rband, \iband) optical photometry of the Type Ia supernova SN 2025bvm, obtained with the COLIBRI telescope at OAN-SPM. The light curves show a photometric decline of $\Delta m_{15}(B) = 0.867 \pm 0.051$~mag, characteristic of a slow-declining Type Ia supernova. After correcting for host galaxy extinction ($E(B-V)_{host} = 0.308 \pm 0.030$~mag) and adopting a distance of 70~Mpc, we derive a peak absolute magnitude of $M_B = -19.13 \pm 0.40$~mag. This luminosity is fully consistent with its slow decline rate, placing SN 2025bvm  within the population of normal Type Ia supernovae. We conclude that SN 2025bvm is a normal Type Ia supernova, whose photometric properties, such as a slow late-time decline and a prominent \iband-band secondary maximum, suggest an explosion that resulted in a particularly massive ejecta.
\end{abstract}

\keywords{NGC 4156, SN 2025bvm,  SN Type Ia, Photometric}

\begin{resumen}
 Presentamos 121 d\'ias de fotometr\'ia \'optica multibanda (\Bband, \gband, \rband, \iband) de la supernova de tipo Ia SN 2025bvm, obtenida con el telescopio COLIBRI del OAN-SPM. Las curvas de luz muestran un declive fotom\'etrico de $\Delta m_{15}(B) = 0.867 \pm 0.051$~mag, caracter\'istico de una supernova de tipo Ia de declive lento. Tras una correcci\'on por la extinci\'on en la galaxia anfitriona ($E(B-V)_{host} = 0.308 \pm 0.030$~mag) y adoptando una distancia de 70~Mpc, derivamos una magnitud absoluta en el pico de $M_B = -19.13 \pm 0.40$~mag. Este valor de luminosidad es plenamente consistente con su lenta tasa de declive, situando a SN~2025bvm  dentro de la poblaci\'on de supernovas de tipo~Ia normales. Concluimos que SN~2025bvm es una supernova de tipo~Ia normal, cuyas propiedades fotom\'etricas, como un declive tard\'io lento y un pronunciado m\'aximo secundario en la banda~\iband, sugieren una explosi\'on que result\'o en una eyecci\'on particularmente masiva.
\end{resumen}

\begin{document}

\maketitle
\pagestyle{fancy}\thispagestyle{firststyle}


\section{INTRODUCTION}

Type Ia supernovae (SNe~Ia) are fundamental tools in modern astrophysics. Their exceptional and consistent peak luminosity have established them as powerful extragalactic distance indicators. In fact, observations of high-redshift SNe Ia provided  key evidence for the  accelerated expansion of the Universe \citep{Riess1998, Perlmutter1999}. This fundamental role as standard candles is based on the remarkable uniformity of their intrinsic photometric properties, allowing for the determination of cosmic distances on cosmological scales

Extensive observational studies of large photometric and spectroscopic samples \citep[e.g.,][]{Hamuy1996, Jha2006} have revealed notable diversity in the observable characteristics of SNe~Ia. Consequently, their use as standard candles requires the application of empirical relations, such as the correlation between peak luminosity and light-curve shape \citep{Phillips1993}, for their calibration and classification. This diversity is a critical aspect, as it offers invaluable clues about the underlying explosion mechanisms and the nature of their progenitor systems \citep[e.g.,][]{Burgaz2025, Rigault2025, Ginolin2025}. Currently, two possible progenitor scenarios are under investigation: the first considers the case of a single-degenerate star (a white dwarf accreting material from a non-degenerate companion), and the second option corresponds to a double-degenerate system (the merger of two white dwarfs). Understanding the wide range of observed light curves is essential for imposing constraints on these explosion models. This diversity has led to the identification of distinct subclasses, including the overluminous and slowly declining SN~1991T-like objects \citep{Filippenko1992a}, the subluminous and rapidly declining SN~1991bg-like objects \citep{Filippenko1992b}, and transitional SNe~Ia that lie between these two categories. In large, unbiased samples, such as that from the ZTF SNIa DR2 \citep{Rigault2025}, normal SNe~Ia form the dominant group, accounting for approximately 70\% of the population. Subluminous (SN 1991bg-like) and overluminous (SN 1991T-like) events comprise roughly 15\% and 5--10\% respectively, with a smaller fraction classified as peculiar \citep{Burgaz2025}. Crucially, this photometric diversity is intrinsically linked to spectroscopic differences. For instance, overluminous events like SN~1991T exhibit hotter spectra near maximum light, characterized by weak or absent Si~II lines but prominent Fe~III features. In contrast, subluminous events, such as SN~1991bg, show cooler spectra with deep Ti~II absorption troughs. While transitional events have been of particular interest of their possible disconnect between photometric and spectroscopic properties \citep{Branch2009}, most SNe~Ia fall into the "normal" category, which are the most widely used as standard candles. In this context, high-cadence, multi-band photometric monitoring of nearby SNe~Ia is essential, as it allows for the precise measurement of decline rates, color indices, and the presence of secondary features, such as the second maximum in the near-infrared, which are key diagnostics for the explosion physics and ejecta properties. A precise characterization of these parameters is fundamental for accurately classifying a Type Ia supernova within all currently known subclasses.

In this study, we present an extensive photometric study of the Type Ia supernova SN~2025bvm. The object was first detected on February 17, 2025, by the ATLAS survey \citep{2025TNSTR.641....1T} in the host galaxy NGC~4156 and was spectroscopically classified as a Type Ia supernova on 2025 February 21 \citep{Balcon2025}. Spectra obtained approximately one week before maximum light, available on the Transient Name Server (TNS), clearly show characteristic Si~II lines and the S~II "W-trough" feature. These spectral characteristics are highly consistent with those of normal Type Ia supernovae, showing a strong resemblance to archetypal events like SN~2011fe. Following its discovery and classification, we performed an intensive multi-band follow-up campaign with the COLIBRI robotic telescope at the Observatorio Astronómico Nacional in San Pedro Mártir, Mexico. The high-cadence data obtained allowed for a robust characterization of its photometric properties, which, as will be demonstrated, confirm its classification as a normal Type Ia supernova, consistent with the general population of these events.

The remainder of this paper is organized as follows. In \S~\ref{sec:obs}, we describe the photometric observations and the data processing. In \S~\ref{sec:analysis}, we present a detailed analysis of the light curves, including a robust methodology for reddening correction and the determination of intrinsic physical properties of the sources. Finally, in \S~\ref{sec:discussion}, we discuss the nature of SN~2025bvm in the context of normal Type Ia supernovae and summarize our results.

\section{Observations and Data Reduction}
\label{sec:obs}

Photometric data were obtained using the 1.3-m COLIBRI  robotic telescope and  DDRAGO instrument \citep{2022SPIE12182E..1SB, 2024SPIE13096E..3DL}, located at the Observatorio Astronómico Nacional de San Pedro Mártir (OAN-SPM), Baja California, Mexico. The observations were made as part of the San Pedro M\'artir Reverberation Mapping program, which primarily monitored the active galactic nucleus (AGN) NGC 4151. As a result of the field coverage, the host galaxy of SN 2025bvm,  NGC 4156, also fell within the same field of view
SN 2025bvm was discovered by the Zwicky Transient Facility (ZTF; \citealt{Bellm2019}) on February 20, 2025, at coordinates RA = 12:10:48.75, DEC = +39:28:28.0 (J2000.0). The Colibri telescope monitored approximately 121 days, from February 23 to June 24, 2025, obtaining photometric data for 79 nights in the \Bband, \gband, \rband, and \iband bands.

\begin{figure*}[!t]
  \centering
  \includegraphics[width=\textwidth]{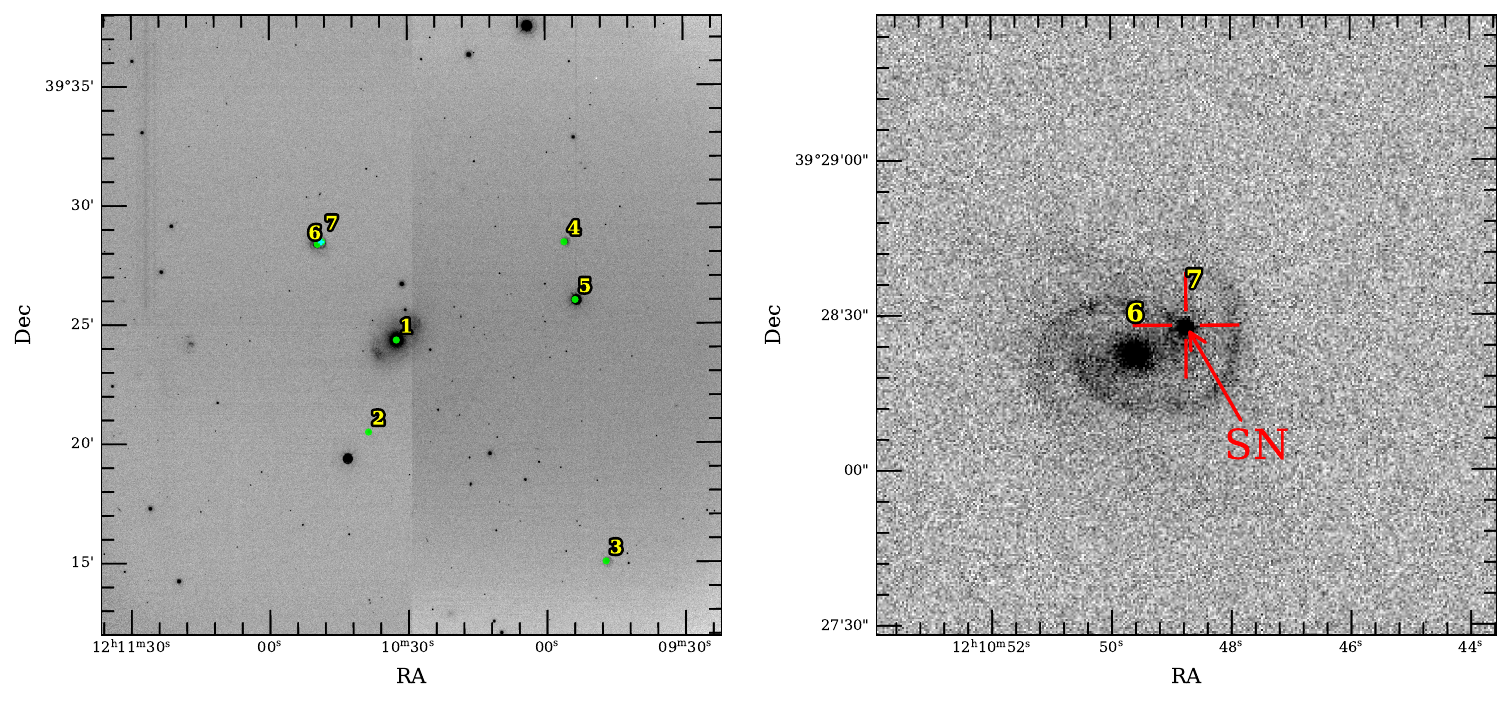}
  \caption{Field of view from the COLIBRI telescope observations. 
  \textbf{Left:} The full field, identifying supernova SN~2025bvm (7) in its host galaxy NGC~4156 (6), the nearby galaxy NGC~4151 (1), and the standard stars used for photometric calibration (listed in Table~\ref{tab:standarStars}). 
  \textbf{Right:} A $5'\times5'$ cutout of the \rband-band image, centered on the host galaxy. The position of SN~2025bvm is marked with red crosshairs. In both panels, North is up and East is to the left.}
  \label{fig:SN}
\end{figure*}

All images obtained in these observations were processed using standard procedures, including bias subtraction and flat-field correction. This data reduction was performed using IRAF \citep{tody1986} and Python packages, specifically \texttt{Astropy} \citep{astropy2013, astropy2018, astropy2022}.  Instrumental magnitudes were extracted using differential photometry, which relies on comparing the brightness of the supernova to a set of reference stars within the same field.
Photometric calibration used a set of standard stars from the supernova field, identified in Figure~\ref{fig:SN} and listed in Table~\ref{tab:standarStars}. The primary references used were stars BD+40~2507 (marked as 2) and UI+40-42 (marked as 3). The magnitudes for these standard stars in the corresponding filters were obtained from public catalogs such as the Sloan Digital Sky Survey \citep{Abdurro2022} and SIMBAD astronomical database\footnote{https://vizier.cds.unistra.fr/viz-bin/VizieR-3?-source=II/336/apass9}.

\begin{table*}[!t]\centering
  \setlength{\tabnotewidth}{0.5\columnwidth}
  \tablecols{3}
  \setlength{\tabcolsep}{2.8\tabcolsep}
  \caption{Photometric standards stars in the SN~2025bvm field. } \label{tab:standarStars}
 \begin{tabular}{cccrrrr}
    \toprule
    \# stars & \multicolumn{1}{c}{$RA$} & \multicolumn{1}{c}{$Dec$} & \multicolumn{1}{c}{\Bband} & \multicolumn{1}{c}{\gband} & \multicolumn{1}{c}{\rband} & \multicolumn{1}{c}{\iband} \\
        & \multicolumn{1}{c}{($deg$)} & \multicolumn{1}{c}{($deg$)} & \multicolumn{1}{c}{(mag)} & \multicolumn{1}{c}{(mag)} & \multicolumn{1}{c}{(mag)} & \multicolumn{1}{c}{(mag)}  \\
    \midrule
    1  & 182.635 & 39.405 & 12.18$\pm$0.001 & 11.91$\pm$0.001 & 11.39$\pm$0.004 & 11.39$\pm$0.003 \\
    2  & 182.680 & 39.323 & 10.41$\pm$0.001 &  10.39$\pm$0.001 & 9.68$\pm$0.001 & 9.54$\pm$0.001 \\
    3  & 182.446 & 39.250 & 12.67$\pm$0.001  & 12.25$\pm$0.001 & 11.77$\pm$0.001 & 11.67$\pm$0.001 \\
    4  & 182.483 & 39.473 & 12.19$\pm$0.001  & 11.84$\pm$0.001 & 10.86$\pm$0.001 & 10.57$\pm$0.001 \\
    5  & 182.473 & 39.433 & 9.81$\pm$0.001  & 9.19$\pm$0.001 & 7.76$\pm$0.001 & 7.32$\pm$0.001 \\
    6  & 182.706 & 39.47 & 15.01$\pm$0.001  & 16.57$\pm$0.044 & 15.90$\pm$0.017 & 15.49$\pm$0.024 \\
    \bottomrule
  \end{tabular}
\end{table*}

Figure~\ref{fig:SN} shows the full field of view, indicating the positions of supernova SN~2025bvm (7), its host galaxy NGC~4156 (6), the nearby galaxy NGC~4151 (1), and the  reference stars. A detailed log of all photometric observations is presented in Table~\ref{tab:photometry}.

\begin{table*}
\centering
\caption{Photometric Observations of SN 2025bvm with the COLIBRI Telescope.}
\label{tab:photometry}
\begin{tabular}{l c cc cc cc cc}\hline
 & & \multicolumn{2}{c}{\Bband-band} & \multicolumn{2}{c}{\gband-band} & \multicolumn{2}{c}{\rband-band} & \multicolumn{2}{c}{\iband-band} \\
\cmidrule(lr){3-4} \cmidrule(lr){5-6} \cmidrule(lr){7-8} \cmidrule(lr){9-10}
Date & MJD & Mag & Err & Mag & Err & Mag & Err & Mag & Err \\
\midrule
2025-02-23 & 60729.520 & 16.979 & 0.015 & 16.967 & 0.012 & 16.442 & 0.017 & 16.854 & 0.025 \\
2025-02-24 & 60730.308 & 16.802 & 0.025 & 16.840 & 0.020 & 16.424 & 0.024 & 16.798 & 0.037 \\
2025-02-25 & 60731.303 & 16.724 & 0.024 & 16.748 & 0.019 & 16.317 & 0.023 & 16.723 & 0.036 \\
2025-02-26 & 60732.288 & 16.600 & 0.022 & 16.637 & 0.018 & 16.250 & 0.022 & 16.676 & 0.036 \\
2025-02-27 & 60733.285 & 16.531 & 0.022 & 16.596 & 0.018 & 16.201 & 0.025 & 16.736 & 0.041 \\
2025-02-28 & 60734.301 & 16.465 & 0.019 & 16.502 & 0.016 & 16.138 & 0.020 & 16.679 & 0.035 \\
2025-03-01 & 60735.294 & 16.425 & 0.019 & 16.492 & 0.016 & 16.060 & 0.019 & 16.681 & 0.035 \\
2025-03-02 & 60736.291 & 16.392 & 0.019 & 16.448 & 0.015 & 16.063 & 0.019 & 16.672 & 0.035 \\
2025-03-05 & 60739.269 & 16.395 & 0.020 & 16.423 & 0.016 & 15.982 & 0.018 & 16.620 & 0.034 \\
2025-03-09 & 60743.302 & 16.527 & 0.029 & 16.500 & 0.022 & 16.005 & 0.022 & 16.727 & 0.043 \\
\multicolumn{10}{c}{...} \\
2025-06-14 & 60840.181 & 18.386 & 0.121 & \nodata & \nodata & \nodata & \nodata & \nodata & \nodata \\
2025-06-16 & 60842.199 & 18.815 & 0.131 & 18.555 & 0.087 & 17.681 & 0.077 & 17.726 & 0.087 \\
2025-06-20 & 60846.229 & 18.838 & 0.136 & 18.534 & 0.083 & 17.652 & 0.073 & 17.691 & 0.090 \\
2025-06-22 & 60848.229 & 18.716 & 0.133 & 18.428 & 0.082 & 17.552 & 0.070 & 17.723 & 0.095 \\
2025-06-24 & 60850.220 & 18.721 & 0.161 & 18.490 & 0.102 & 17.626 & 0.088 & 17.724 & 0.114 \\
\bottomrule
\end{tabular}
\end{table*}

\section{Light Curve Analysis and Derived Properties}
\label{sec:analysis}

In this section, we present a comprehensive analysis of the COLIBRI photometric data for SN~2025bvm. We first characterize the observed light curves and their main parameters, then derive the host galaxy extinction based on the color evolution, and finally compute the intrinsic properties of the supernova.


\subsection{Light Curves and Photometric Parameters}
\label{subsec:lightcurves}

\begin{figure}[!t]
  \includegraphics[width=\columnwidth]{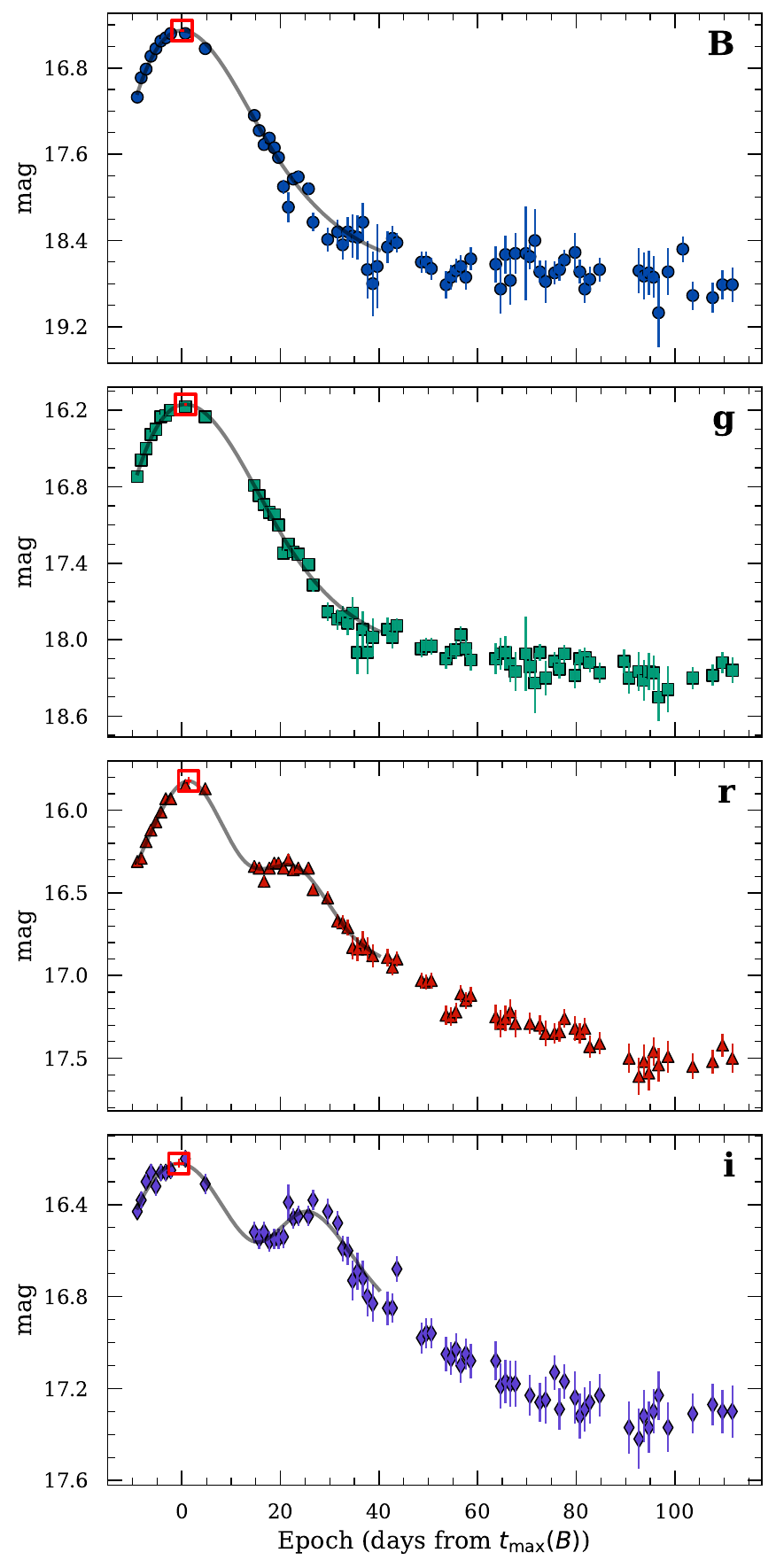}
  \caption{The observed \Bband, \gband, \rband, and \iband-band light curves of SN 2025bvm from the COLIBRI telescope. SNooPy2 model fits to the multi-band light curves of SN~2025bvm to determine the peak parameters. The solid grey line shows the best-fit template model in each band. The red square marks the resulting time ($t_{max}$) and magnitude ($m_{max}$) of the peak.}
  \label{fig:peak_fits}
\end{figure}

The complete multi-band optical light curves of SN~2025bvm are presented in Figure~\ref{fig:peak_fits}. The high-cadence observations provide excellent coverage of the supernova's evolution, from approximately $\sim9$ days before the B-band maximum until more than 110 days after, capturing the rise, peak, and subsequent decline phases in detail.

The shape of the light curves is characteristic of a Type~Ia supernova \citep[e.g.,][]{Filippenko1997}. In particular, as shown in Figure~\ref{fig:peak_fits}, the light curves of SN~2025bvm provide a clear frame of reference by showing distinctive features of normal Type~Ia supernovae. These include a prominent secondary maximum in the \iband-band and an inflection point or "shoulder" in the \rband-band, both occurring approximately 30 days after the B-band maximum. As is typical, the evolution shows a clear wavelength dependence, reaching maximum brightness earlier and declining more steeply in the bluer bands \citep[see, e.g.,][]{Leibundgut1991, Hamuy1996}. The object's brightest observed peak is in the \rband-band, with an apparent magnitude of $m_r \approx 15.8$~mag.

To derive the primary photometric parameters, two packages were used, SNooPy2 (SuperNovea in object-oriented Python)\footnote{https://csp.obs.carnegiescience.edu} \citep{Burns2011, Burns2014} and SALT2\footnote{https://dessn.github.io/sn-doc/doc/out/html/inv\_gauss.html\#inv-gauss} \citep{Guy2007}, the latter implemented through the \texttt{sncosmo} code \citep{Barbary2023}. These methods fit the light curves using empirical models that describe the evolution of the spectral energy distribution (SED) of a Type~Ia supernova. By fitting these models to the multi-band photometric data, a robust and simultaneous estimation of the peak magnitude ($m_\mathrm{max}$), the time of maximum light ($t_\mathrm{max}$), and the light-curve shape parameters is obtained. This approach is particularly effective for handling light curves with temporal gaps, such as the one observed in our data between +7 and +15 days post-maximum. The results from both fitting methods, presented in Table~\ref{tab:peak_params}, are in excellent agreement, providing high confidence in the derived parameters.

\begin{table*}[!t]\centering
  \setlength{\tabnotewidth}{0.5\columnwidth}
  \tablecols{3}
  \setlength{\tabcolsep}{2.8\tabcolsep}
  \caption{Photometric Parameter of SN2025bvm} \label{tab:peak_params}
 \begin{tabular}{llrll}
    \toprule
    Filter & \multicolumn{1}{c}{$t_{max}$ (MJD) } & \multicolumn{1}{c}{$m_{max}$ (mag)} & \multicolumn{1}{c}{$t_{max}$ (MJD) } & \multicolumn{1}{c}{$m_{max}$ (mag)} \\
        & \multicolumn{1}{c}{SNooPy} & \multicolumn{1}{c}{SNooPy} & \multicolumn{1}{c}{SALT2} & \multicolumn{1}{c}{SALT2}  \\
    \midrule
    \Bband  & 60738.56$\pm$0.44 & 16.452$\pm$0.014 & 60738.24$\pm$0.09 & 16.427$\pm$0.041 \\
    \gband  & 60739.28$\pm$0.24 & 16.153$\pm$0.009 & 60738.62$\pm$0.09 &  16.281$\pm$0.036 \\
    \rband  & 60739.91$\pm$0.64 & 15.823$\pm$0.026 & 60741.43$\pm$0.08 & 15.771$\pm$0.039 \\
    \iband  & 60737.93$\pm$1.05 & 16.221$\pm$0.020 & 60734.84$\pm$0.09 & 16.068$\pm$0.048 \\
    \bottomrule
  \end{tabular}
\end{table*}

\begin{figure}[!t]
  \includegraphics[width=\columnwidth]{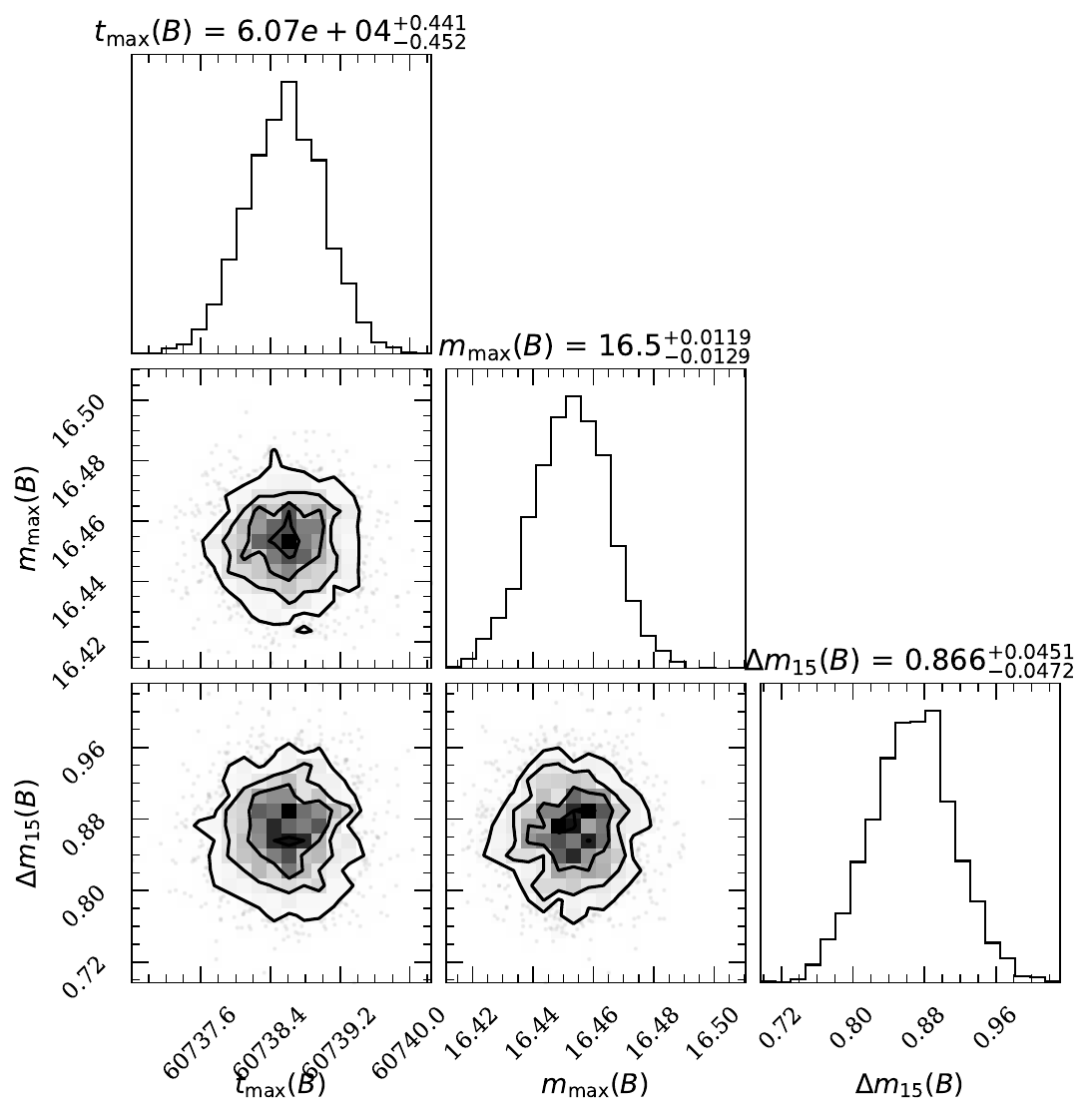}
  \caption{Corner plot showing the posterior probability distributions for the key \Bband-band light-curve parameters of SN~2025bvm, as derived from the SNooPy2 fit. The diagonal panels show the  probability distribution for each parameter: the time of maximum ($t_{max}(B)$), the apparent magnitude at maximum ($m_{max}(B)$), and the decline rate ($\Delta m_{15}(B)$). The off-diagonal panels show the covariances between parameter pairs. The contours represent the 1, 2, and 3-$\sigma$ confidence levels.}
  \label{fig:corner_plot}
\end{figure}

From the SNooPy2 fit, we measured the decline rate parameter for the \Bband-band light curve, $\Delta m_{15}(B)$, which is the magnitude difference between the peak and 15 days post-maximum. We find a value of $\Delta m_{15}(B) = 0.867 \pm 0.051$~mag. This measurement and its covariance with the peak time and magnitude are illustrated in the corner plot in Figure~\ref{fig:corner_plot}. This value places SN~2025bvm firmly within the range of normal Type~Ia supernovae \citep[whose B-band range is between 0.7 and 1.7, according to][]{Burns2014}, confirming that SN~2025bvm is a normal Type~Ia supernova belonging to the slow-declining subgroup.

Furthermore, our high-quality \gband-band observations allow for an additional consistency check. The parameters derived from our \gband-band light curve are fully consistent with the trends and relationships established for the large, homogeneous sample of normal Type~Ia supernovae from the ZTF DR2 \citep[e.g.,][]{Rigault2025, Ginolin2025}. This consistency further strengthens our classification of SN~2025bvm as a bona fide normal Type~Ia supernova.

\subsection{Color Evolution and Host Galaxy Reddening}
\label{subsec:reddening}

To derive the intrinsic properties of SN~2025bvm, it is essential to correct the observed photometry for interstellar dust extinction, both within the Milky Way and in the host galaxy. First, the Galactic component of reddening in the direction of NGC~4156 was estimated to be $E(B-V)_{MW} = 0.0198$~mag, based on the dust maps of \citet{Schlafly2011}.

\begin{figure*}[!t]
  \centering
  \includegraphics[width=1.00\textwidth]{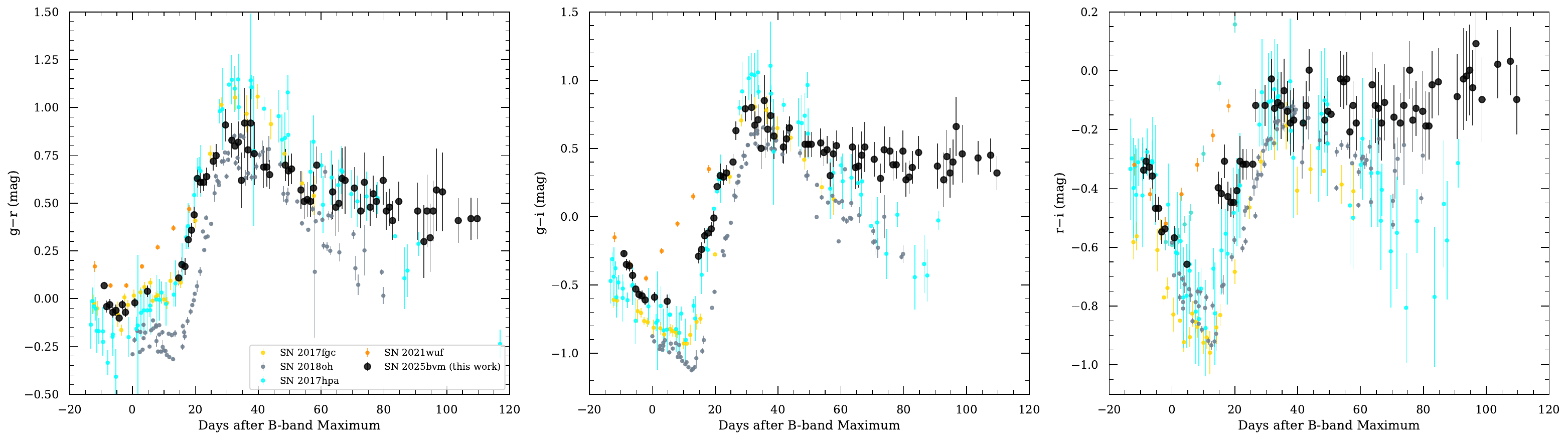}
  \caption{Intrinsic color evolution of SN~2025bvm (black points) compared to a sample of normal Type~Ia supernovae. From left to right, the panels show the evolution of the $(g-r)$, $(g-i)$, and $(r-i)$ colors. All supernovae shown have been corrected for their total line-of-sight extinction (both Galactic and host galaxy). The good morphological agreement of SN~2025bvm with the comparison sample across all three colors supports our derived reddening value and confirms its nature as a normal Type~Ia supernova.}
  \label{fig:color_comparison}
\end{figure*}

To estimate the host galaxy reddening, $E(B-V)_{host}$, the color evolution of SN~2025bvm was compared with that of a sample of well-studied normal Type~Ia supernovae. This method, which uses high-cadence data instead of low-resolution templates, allows for a more precise calibration by capturing the detailed structure of the color curves. The comparison sample includes the events SN~2002bo \citep{Benetti2004}, SN~2017fgc \citep{Zeng2021}, SN~2018oh \citep{Li2019}, SN~2017hpa \citep{Zeng2021}, and SN~2021wuf \citep{Zeng2024}. Figure~\ref{fig:color_comparison} presents the evolution of the $(g-r)$, $(g-i)$, and $(r-i)$ colors of SN~2025bvm, after being corrected only for Galactic reddening, compared to the sample of normal SNe~Ia, which have already been corrected for their respective total reddening.

A clear vertical offset is observed between the color curves of SN~2025bvm and those of the comparison sample, indicating significant reddening in the host galaxy. A reddening value of $E(B-V)_{host} = 0.308 \pm 0.030$~mag was determined to best align all three color curves of SN~2025bvm simultaneously with the average behavior of the normal SNe~Ia sample, as shown in Figure~\ref{fig:color_comparison}. For this derivation, a standard extinction law with $R_V = 3.1$ \citep{Cardelli1989} was assumed. Although the value of $R_V$ is known to vary along different lines of sight \citep[e.g.,][]{Szabo2003,Burns2014}, the consistency obtained across multiple color curves, including the $(r-i)$ color which is less sensitive to extinction, supports the validity of this assumption in our case. Furthermore, while the synthesized Nickel mass can affect intrinsic colors \citep{Konyves-Toth2020}, the excellent morphological agreement with the sample of normal supernovae suggests that SN~2025bvm does not exhibit intrinsic color peculiarities that would significantly affect our reddening calculation. Summing both components, the total line-of-sight reddening is $E(B-V)_{total} = 0.328 \pm 0.030$~mag.


\subsection{Corrected Light Curves and Absolute Magnitude}
\label{subsec:corrected_lcs}

\begin{figure}[!t]
  \includegraphics[width=\columnwidth]{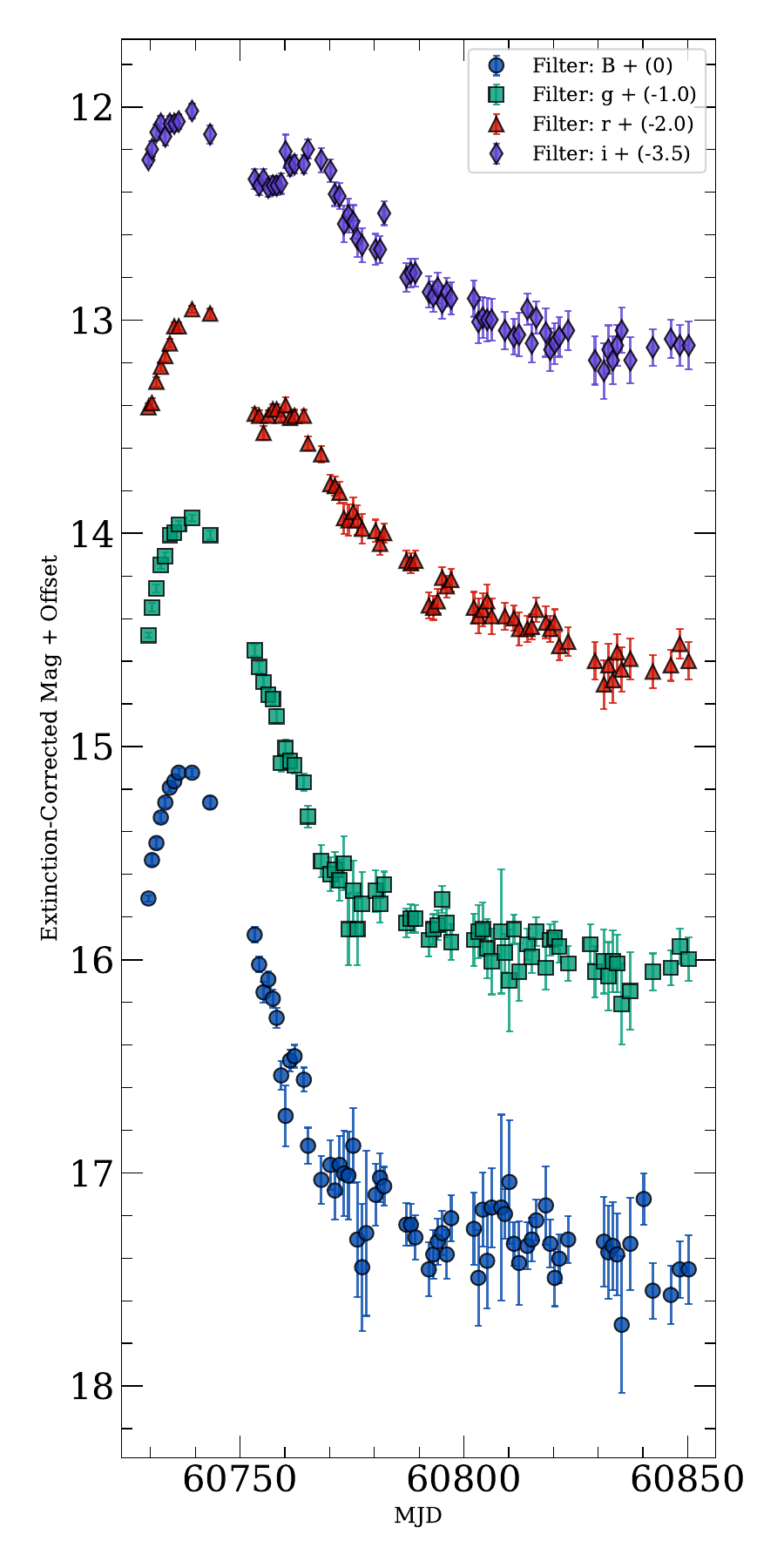}
  \caption{The intrinsic light curves of SN~2025bvm after applying the full extinction correction. For visualization purposes, the light curves have been shifted vertically by the offsets listed in the legend. The correction reveals the intrinsically blue nature of the supernova at maximum light, with the \Bband-band peak being significantly brighter than in the redder bands.}
  \label{fig:lcs_corrected}
\end{figure}

Applying the total line-of-sight extinction of $E(B-V)_{total} = 0.328$~mag, derived in the previous section, we obtain the final intrinsic light curves of SN~2025bvm. These extinction-corrected light curves are presented in Figure~\ref{fig:lcs_corrected}.

The correction is strongly wavelength-dependent, having a much greater impact on the bluer bands. The \Bband-band peak, for instance, is brightened by approximately 1.55~mag, whereas the \iband-band peak is brightened by only 0.78~mag. While the observed photometry showed the peaks of all filters clustered in magnitude, Figure~\ref{fig:lcs_corrected} clearly shows that SN~2025bvm was intrinsically a very blue object at its brightest, with the \Bband-band peak now significantly brighter than the \rband-band and \iband-band peaks.
With the fully corrected photometry, we derive the peak absolute magnitude of the supernova. Adopting a distance modulus to the host galaxy NGC~4156 of $\mu = 34.21 \pm 0.40$~mag (corresponding to a distance of $70 \pm 13$~Mpc; \citep[e.g.,][]{Nieto1984, Stevance2025}), and using the B-band maximum magnitude from the SNooPy fit ($m_{max} = 16.452 \pm 0.014$~mag) to derive a corrected apparent magnitude at the B-band peak of $m_{B,corr} = 15.08$~mag, we obtain an absolute magnitude of:
\textbf{\[ M_B = m_{B,corr} - \mu = 15.08 - 34.21 = -19.13 \pm 0.40  mag \]}
This value is fully consistent with the luminosity of a normal Type~Ia supernova. Typical Type~Ia supernovae cluster around a peak absolute magnitude of $M_B \approx -19.25$~mag \citep[e.g.,][]{Richardson2002, Taubenberger2017}. Therefore, both its light-curve shape and its intrinsic brightness place SN~2025bvm firmly within the population of normal Type~Ia supernovae.

A summary of the reddening parameters, decline rate, and derived physical properties for SN 2025bvm is presented in Table~\ref{tab:parameter_ext}.

\begin{table}[!t]\centering
  \caption{Derived Parameters of SN2025bvm} 
  \label{tab:parameter_ext}
 \begin{tabular}{ll}
    \toprule
    Parameter & \multicolumn{1}{c}{Value (mag)}\\
    \midrule
    $E(B-V)_{MW}$    & 0.0198 (SF11) \\
    $E(B-V)_{host}$  & 0.308$\pm$0.030 \\
    $E(B-V)_{total}$ & 0.328 $\pm$0.030 \\
    \addlinespace 
    Decline Rate $\Delta m_{15} (B)$  & 0.867$\pm$0.051 \\
    Classification & Normal SN \\
    \addlinespace 
    Absolute Magnitude $M_B$ & $-19.13$  (d = 70 Mpc) \\
    \bottomrule
  \end{tabular}
\end{table}


\subsection{Distance and Quasi-bolometric Light Curve}
\label{subsec:bolometric}

To determine the physical properties of SN~2025bvm, it is crucial to establish a  distance to its host galaxy, NGC~4156. Using the galaxy's redshift, $z=0.0163$, and assuming a standard cosmological model ($\text{H}_0 = 70$~km~s$^{-1}$~Mpc$^{-1}$, $\Omega_M = 0.3$, $\Omega_\Lambda = 0.7$), a luminosity distance of 73.12~Mpc is obtained. On the other hand, an average of independent measurements from the literature yields a distance modulus of $\mu = 34.21 \pm 0.40$~mag \citep[e.g.,][]{Nieto1984, Stevance2025}, which corresponds to a distance of $70 \pm 13$~Mpc. Given the consistency between both methods, in this work we adopt a distance of $d = 70 \pm 13$~Mpc for all subsequent calculations.
With this distance, the absolute peak magnitude in the \Bband\ for SN~2025bvm is $M_B = -19.13 \pm 0.40$~mag. This value, combined with the decline rate of $\Delta m_{15}(B) = 0.867 \pm 0.051$~mag, places SN~2025bvm in the region of normal, luminous Type~Ia supernovae, consistent with the width-luminosity relation \citep{Phillips1993}.

\begin{figure}[!t]
  \includegraphics[width=\columnwidth]{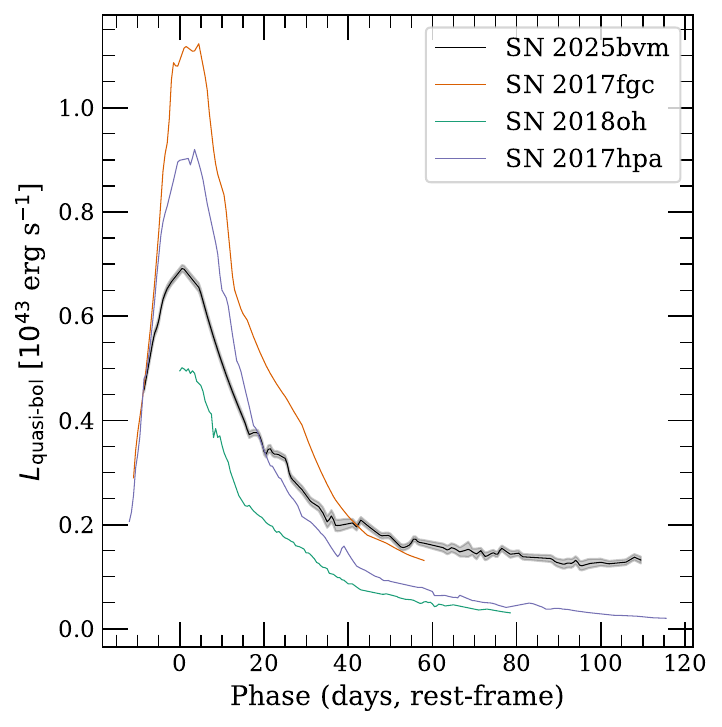}
  \caption{Comparison of the quasi-bolometric light curves. The light curve of SN~2025bvm is shown (black line) with its 1-$\sigma$ error band shaded, alongside a sample of other normal Type~Ia supernovae. The figure places SN~2025bvm in an intermediate luminosity range within the normal class, being less luminous than the high-luminosity event SN~2017fgc but more luminous than SN~2018oh. This confirms its classification as a normal Type~Ia supernova.}
  \label{fig:lc_bol}
\end{figure}

Additionally, the quasi-bolometric light curve of SN~2025bvm was constructed. To do this, the extinction-corrected fluxes in the \Bband, \gband, \rband, and \iband\ bands were integrated. An empirical correction was applied to this optical flux to estimate the contribution from the ultraviolet (UV) and near-infrared (NIR) bands, assuming that the flux fraction in these regions is similar to that observed in well-studied normal Type~Ia supernovae \citep[e.g., SN~2011fe;][]{Burns2018}. Figure~\ref{fig:lc_bol} shows the resulting quasi-bolometric light curve for SN~2025bvm, compared to those of other normal Type~Ia supernovae. SN~2025bvm reached a peak bolometric luminosity of $L_{peak} = (6.91 \pm 0.66) \times 10^{42}$~erg~s$^{-1}$. This value is comparable to that of other normal SNe~Ia, although slightly less luminous than high-luminosity events like SN~2017fgc ($L_{peak} \approx 1.12 \times 10^{43}$~erg~s$^{-1}$).

The mass of $^{56}$Ni synthesized during the explosion can be estimated from the peak luminosity using Arnett's Law \citep{Arnett1982}, which relates the peak of the bolometric light curve to the amount of radioactive material. The relation is expressed as:

\begin{equation}
    L_{peak} \approx \left( 6.45e^{-t_r / 8.8\text{d}} + 1.45e^{-t_r / 111.3\text{d}} \right) \frac{M_{Ni}}{M_{\odot}} \times 10^{43} \text{ erg s}^{-1}
    \label{eq:arnett}
\end{equation}

where $t_r$ is the rise time of the bolometric light curve and $M_{Ni}$ is the Nickel mass in solar masses. Assuming a rise time of $t_r \approx 18.0$~days \citep[e.g.,][]{Riess1998, Firth2015}, the peak luminosity of SN~2025bvm corresponds to a Nickel mass of $M_{Ni} \approx 0.34~M_{\odot}$. This $M_{Ni}$ value is consistent with that expected for a Type~Ia supernova of normal luminosity, which reinforces our overall classification of the event.
\subsection{Late-Time Decline and i-band Secondary Maximum}
\label{subsec:late_time}

\begin{figure*}[!t]
\centering
  \includegraphics[width=0.80\linewidth]{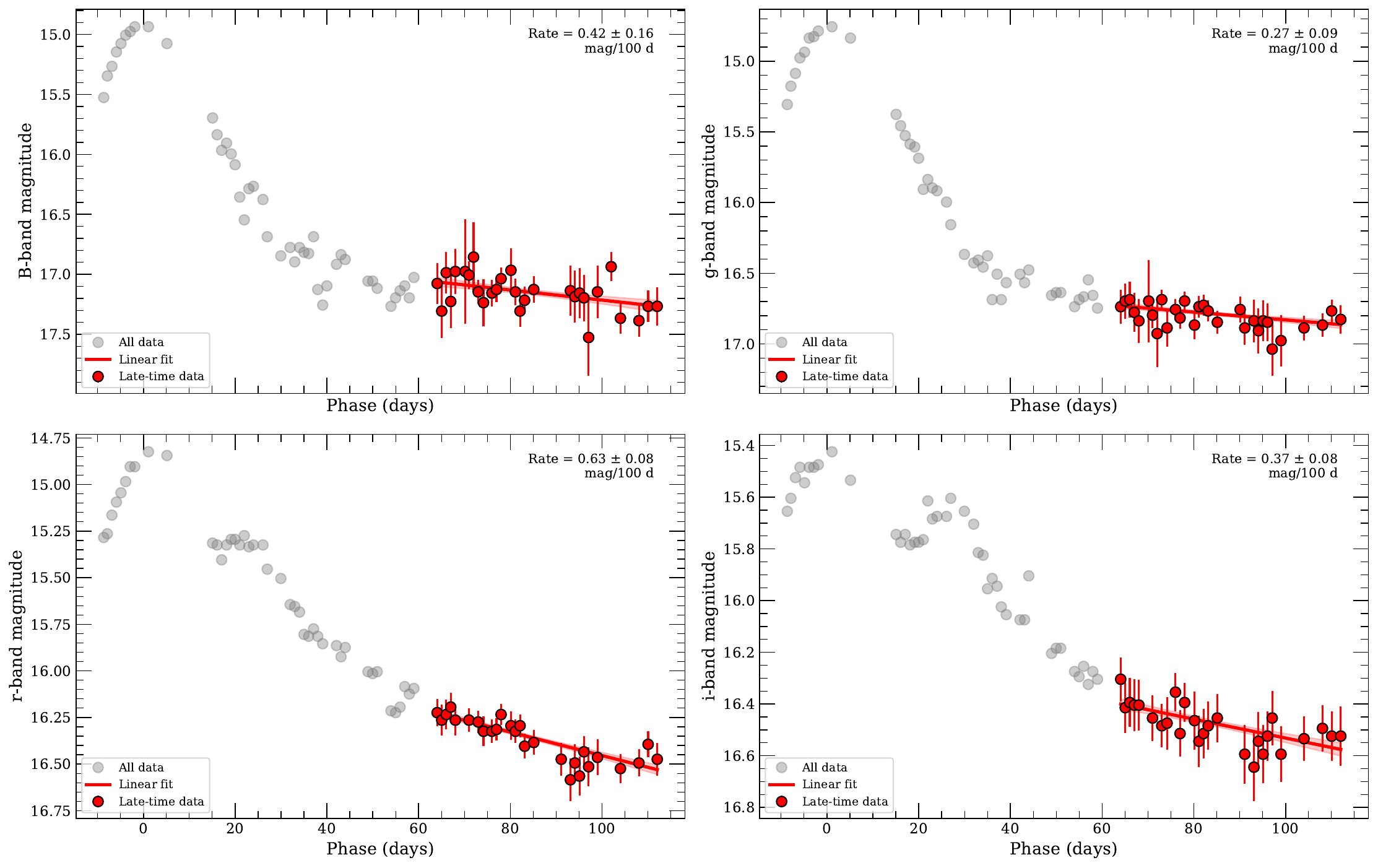}%
  \caption{Linear fits to the intrinsic light curves of SN~2025bvm in the late-time phase (phase > +60 days). In each panel, the full light curve is shown with gray points for context. The red points are the data used in the linear fit. The solid red line represents the best-fit model, while the shaded band indicates the 1-$\sigma$ confidence interval of the fit. The resulting decline rate, with its uncertainty, is shown in units of magnitudes per 100 days for each filter.}
  \label{fig:late_time_fits}
\end{figure*}

The late-time luminosity of a Type Ia supernova, typically at phases beyond +60 days, is powered by the radioactive decay of $^{56}$Co into $^{56}$Fe \citep[e.g.,][]{Weaver1980, Colgate1980}. The rate at which the light curve declines in this phase is a direct probe of the efficiency with which gamma-rays from this decay are trapped and thermalized within the expanding ejecta. A slower decline rate implies a more massive or denser ejecta that is more effective at trapping this high-energy radiation \citep{Sollerman2004}. We measured the decline rates by performing a linear fit to the corrected light curves in each band for all data points with a phase greater than +60 days. The uncertainties on the decline rates were estimated using a Monte Carlo approach, taking into account both the photometric errors and the data scatter. The fits are presented in Figure~\ref{fig:late_time_fits}, and the resulting rates are summarized in Table~\ref{tab:late_time_rates}.

We measure slow decline rates in all bands; for example, $0.42 \pm 0.16$ mag/100 days in the \Bband-band and $0.37 \pm 0.08$ mag/100 days in the \iband-band. Although our observations in this phase are noisy, these rates are significantly slower than those of typical SNe~Ia, which decline at approximately $1.4 - 1.6$ mag/100 days in the \Bband-band at similar phases \citep{Lira1998}, and are comparable to or even slower than those observed in luminous 91T-like events. Such a slow decline provides strong evidence for a very massive and dense ejecta with highly efficient gamma-ray trapping.

\begin{table}[!t]\centering
  \caption{Late-Time Decay Rates} 
  \label{tab:late_time_rates}
 \begin{tabular}{lc}
    \toprule
    Filter & \multicolumn{1}{c}{Decay Rate (mag/100 days)}\\
    \midrule
    \Bband  & 0.42 $\pm$ 0.16 \\
    \gband  & 0.27 $\pm$ 0.09 \\
    \rband  & 0.63 $\pm$ 0.08 \\
    \iband  & 0.37 $\pm$ 0.08 \\
    \bottomrule
  \end{tabular}
\end{table}

Another powerful diagnostic, particularly sensitive to the ejecta's opacity and thermal structure, is the shape of the \iband-band light curve. Luminous SNe~Ia, such as SN~1991T, typically exhibit a prominent secondary maximum or a distinct shoulder around 30 days past the \Bband-band peak. This feature is attributed to a temporary decrease in opacity in the near-infrared as the ejecta cools and recombines \citep[e.g.,][]{Kasen2006}. In contrast, subluminous events like SN~1991bg show a monotonic decline with no secondary peak \citep{Leibundgut1991, Taubenberger2008}. In Figure~\ref{fig:i_band_comparison}, we compare the shape of the \iband-band light curve of SN~2025bvm to those of different SNe~Ia subclasses, after normalizing all curves to their peak brightness. The figure clearly shows that SN~2025bvm exhibits a strong and well-defined secondary maximum, the behavior of which is highly consistent with that of normal (like SN~2011fe) and luminous (like SN~1991T) SNe~Ia.

\begin{figure}[!t]
  \includegraphics[width=\columnwidth]{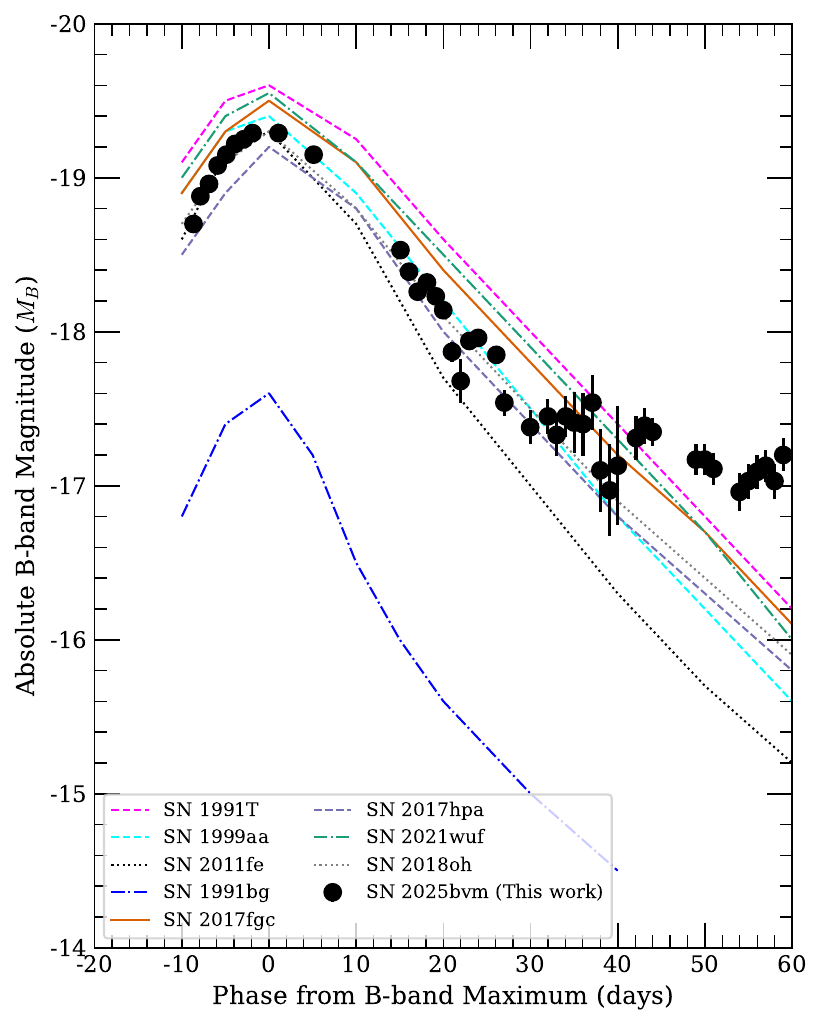}%
  \caption{The absolute \Bband-band light curve of SN~2025bvm (black points) compared to a sample of SNe~Ia  representing different subclasses. After correcting for distance and reddening, the luminosity of SN~2025bvm is fully consistent with that of a normal Type~Ia supernova (like SN~2011fe) and is clearly distinct from subluminous (SN~1991bg) and overluminous (SN~1991T) events.}
  \label{fig:B_band_comparison}
\end{figure}

Both the slow late-time decline and the prominent \iband-band secondary maximum provide two independent lines of photometric evidence suggesting that SN~2025bvm possesses a massive ejecta, characteristic of a high-energy event. These properties, combined with its normal SN~Ia intrinsic luminosity, point to its classification as a normal Type~Ia event, but on the high-mass, slow-declining end of the distribution.

\section{Discussion and Conclusions}
\label{sec:discussion}

Our high-cadence photometric analysis of SN~2025bvm has allowed for a robust characterization of its physical properties. The parameters derived in this work, summarized in Table~\ref{tab:summary_params}, are a decline rate of $\Delta m_{15}(B) = 0.867 \pm 0.051$~mag and a peak absolute magnitude of $M_B = -19.13 \pm 0.40$~mag. These results place SN~2025bvm within the population of normal Type~Ia supernovae.

Figure~\ref{fig:B_band_comparison} illustrates this point, showing the absolute B-band light curve of SN~2025bvm along with those of several archetypal SNe~Ia. As observed, the curve of SN~2025bvm aligns very well with that of a normal SN~Ia like SN~2011fe and is clearly distinct from subluminous (SN~1991bg) and overluminous (SN~1991T) events. This consistency is reinforced when comparing SN~2025bvm with a broader sample of well-studied events. With a luminosity of $M_B \approx -19.13$~mag and a $\Delta m_{15}(B) \approx 0.87$, SN~2025bvm is located in the high-luminosity, slow-declining region of the Phillips diagram. It is slightly less luminous than events with a similar decline rate, such as SN~2017fgc ($\Delta m_{15}(B) \sim 0.75$, $M_B \sim -19.5$~mag) or SN~1991T ($\Delta m_{15}(B) \sim 0.65$, $M_B \sim -19.6$~mag), but significantly more luminous than intermediate-decline-rate SNe~Ia like SN~2018oh ($\Delta m_{15}(B) \sim 1.1$, $M_B \sim -19.2$~mag). This positioning is fully consistent with the diversity observed in large samples like the ZTF SNIa DR2, which shows a continuum of properties rather than discrete classes \citep{Burgaz2025, Rigault2025}.

\begin{figure}[!t]
  \includegraphics[width=\columnwidth]{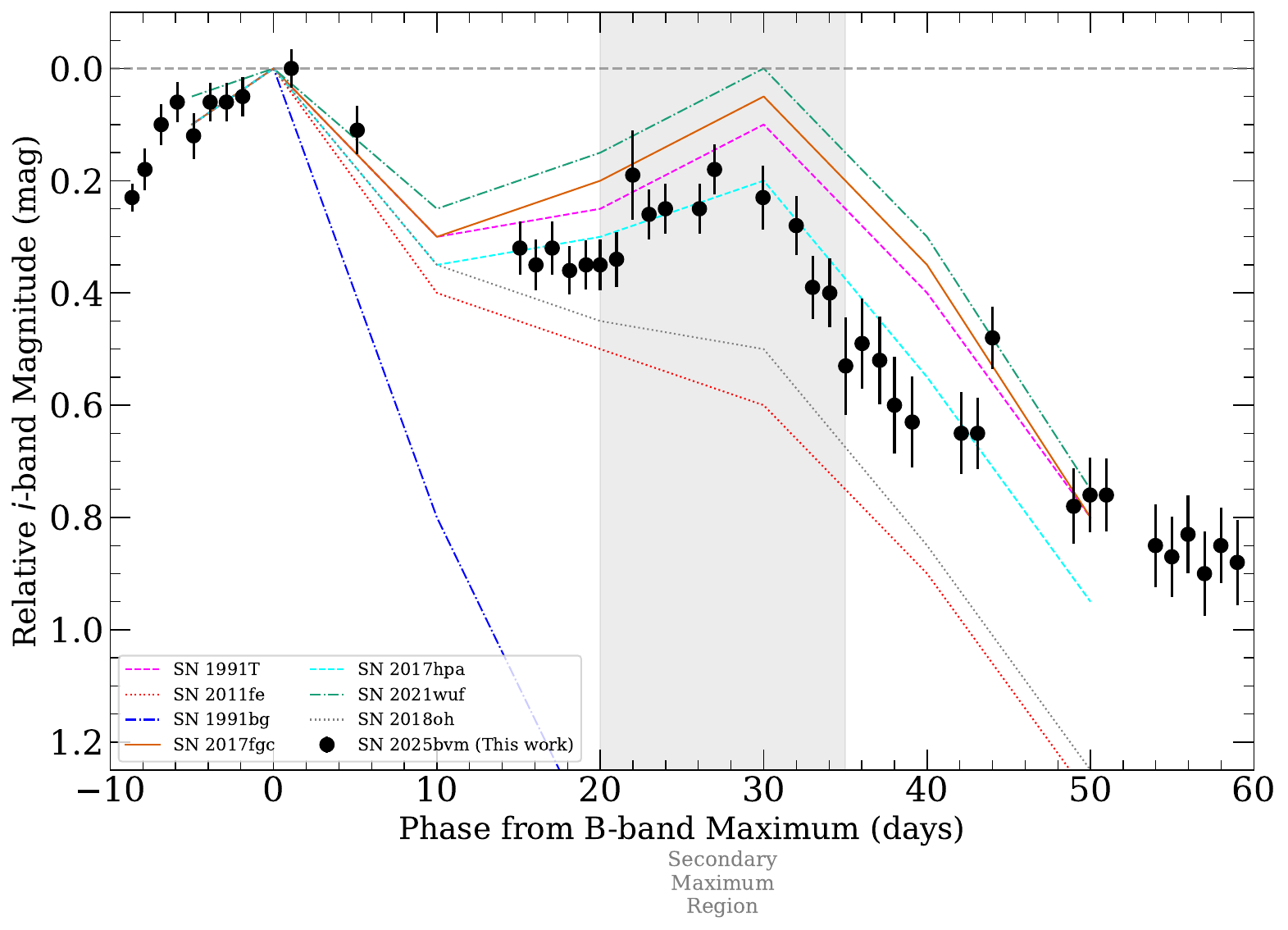}
  \caption{The normalized \iband-band light curve of SN~2025bvm (black points) compared to a sample of templates from different SNe~Ia subclasses. All curves have been normalized to their peak brightness. The presence of a strong secondary maximum in the shaded region (around 30 days past the B-band maximum) is a key feature. The shape of SN~2025bvm's light curve is consistent with that of normal and luminous SNe~Ia, and clearly differs from the monotonic decline of the subluminous 91bg subclass.}
  \label{fig:i_band_comparison}
\end{figure}

From a physical standpoint, the peak quasi-bolometric luminosity of $L_{peak} = (6.91 \pm 0.66) \times 10^{42}$~erg~s$^{-1}$ corresponds to a synthesized Nickel mass of $M_{Ni} \approx 0.34~M_{\odot}$. This value, while typical for SNe~Ia, is relatively modest for an event with such a broad light curve. However, the morphological features, such as the slow late-time decline (e.g., $0.42 \pm 0.16$~mag/100 days in the B-band) and the strong secondary maximum in the i-band (Figure~\ref{fig:i_band_comparison}), are indicators of an ejecta with a relatively high mass and opacity. This suggests that SN~2025bvm, while a normal SN~Ia in terms of luminosity, is located at the high-ejecta-mass end of the distribution of normal events, which explains its high efficiency in trapping gamma-rays in late phases.

\begin{table*}[!t]\centering
  \caption{Summary of Derived Parameters for SN~2025bvm}
  \label{tab:summary_params}
  \begin{tabular}{ll}
    \toprule
    Parameter & \multicolumn{1}{c}{Value} \\
    \midrule
    \multicolumn{2}{c}{\textit{Distance and Reddening Parameters}} \\
    \addlinespace
    Redshift, $z$ & 0.0163 \\
    Distance, $d$ (Mpc) & $70 \pm 13$ \\
    Distance Modulus, $\mu$ (mag) & $34.21 \pm 0.40$ \\
    Galactic Reddening, $E(B-V)_{MW}$ (mag) & 0.0198 \\
    Host Reddening, $E(B-V)_{host}$ (mag) & $0.308 \pm 0.030$ \\
    Total Reddening, $E(B-V)_{total}$ (mag) & $0.328 \pm 0.030$ \\
    \addlinespace
    \multicolumn{2}{c}{\textit{Photometric and Physical Parameters}} \\
    \addlinespace
    B-band Peak, $t_{max}(B)$ (MJD) & $60738.56 \pm 0.44$ \\
    Decline Rate, $\Delta m_{15}(B)$ (mag) & $0.867 \pm 0.051$ \\
    Absolute Magnitude, $M_B$ (mag) & $-19.13 \pm 0.40$ \\
    Peak Bolometric Luminosity, $L_{peak}$ ($10^{42}$ erg s$^{-1}$) & $6.91 \pm 0.66$ \\
    $^{56}$Ni Mass, $M_{Ni}$ ($M_{\odot}$) & $\approx 0.34$ \\
    Photometric Classification & Normal Ia (slow-declining subgroup) \\
    \bottomrule
  \end{tabular}
\end{table*}

The main conclusions of this work can be summarized as follows:
\begin{enumerate}
    \item We have presented high-cadence optical photometry of SN~2025bvm in the \Bband, \gband, \rband, and \iband\ bands, covering the rise phase until more than 100 days after maximum light.
    \item Using standard light-curve fitting tools (SNooPy2 and SALT2), the photometric parameters were robustly derived, yielding $\Delta m_{15}(B) = 0.867 \pm 0.051$~mag.
    \item An estimate of the host galaxy reddening ($E(B-V)_{host} = 0.308 \pm 0.030$~mag) was performed and a distance consistent with the literature ($d = 70$~Mpc) was adopted, leading to an absolute magnitude of $M_B = -19.13 \pm 0.40$~mag.
    \item The quasi-bolometric light curve was constructed, yielding a peak luminosity of $L_{peak} = (6.91 \pm 0.66) \times 10^{42}$~erg~s$^{-1}$, which corresponds to a $^{56}$Ni mass of $\approx 0.34~M_{\odot}$.
    \item We conclude that SN~2025bvm is a normal, luminous Type~Ia supernova, whose set of properties (luminosity, light-curve shape, and late-time decline) point to an explosion that produced a moderate Nickel mass within a particularly massive ejecta.
\end{enumerate}

\renewcommand{\refname}{REFERENCES}
\bibliography{rmaa}


\section{ACKNOWLEDGEMENTS}
D.H.G-B. acknowledges support from the ‘Investigadores e Investigadoras por México’ fellowship program of the Secretaría de Ciencia, Humanidades, Tecnología e Inovación (SECIHTI) of Mexico, which made this research possible at the Instituto de Astronomía, Universidad Nacional Autónoma de México (IA-UNAM), Ensenada campus. Ma.T.G-D gratefully acknowledge support from the PAPIT through grant AG101223. The data used in this paper were acquired with the DDRAGO instrument on the COLIBRÍ telescope at the Observatorio Astronómico Nacional on the Sierra de San Pedro Mártir. COLIBRÍ and DDRAGO are funded by the Universidad Nacional Autónoma de México (CIC and DGAPA/PAPIIT IN109418 and IN109224), and CONAHCyT (1046632 and 277901). COLIBRI received financial support from the French government under the France 2030 investment plan, as part of the Initiative d’Excellence d’Aix-Marseille Université-A*MIDEX (ANR-11-LABX-0060 – OCEVU and AMX-19-IET-008 – IPhU), from LabEx FOCUS (ANR-11-LABX-0013), from the CSAA-INSU-CNRS support program, and from the International Research Program ERIDANUS from CNRS. COLIBRÍ and DDRAGO are operated and maintained by the Observatorio Astronómico Nacional and the Instituto de Astronomía of the Universidad Nacional Autónoma de México.



\end{document}